\documentclass[11pt,twoside]{article}
\usepackage{asp2010}

\begin{document}

\title{Searching for variable stars in the VVV globular clusters}
\author{Javier~Alonso-Garc\'ia,$^{1,2}$ Istv\'an~D\'ek\'any,$^{1,2}$ M\'arcio~Catelan,$^{1,2}$ Rodrigo~Contreras~Ramos,$^{1,2}$, and Dante~Minniti$^{1,2,3}$
\affil{$^1$Instituto de Astrof\'isica, Facultad de F\'isica, Pontificia Universidad Cat\'olica de Chile, Av. Vicu\~na Mackenna 4860, 782-0436 Macul, Santiago, Chile }
\affil{$^2$The Milky Way Millennium Nucleus, Av. Vicu\~{n}a Mackenna 4860, 
782-0436 Macul, Santiago, Chile}
\affil{$^3$Vatican Observatory, Vatican City State V-00120, Italy}}

\begin{abstract}
  The VVV is currently surveying the central regions of the Milky Way
  at near-infrared wavelengths, including 36 known Galactic globular
  clusters. We already have data in several filters to generate their
  color-magnitude diagrams. We also have enough epochs to begin
  producing the light-curves of the cluster stars and look for any
  possible member variable stars. We are especially interested in
  RR~Lyrae stars, since they are abundant in Galactic globular
  clusters, and the period-luminosity relation they show at
  near-infrared wavelengths, can help enormously in providing accurate
  distances and reddenings for the most extincted and poorly studied
  inner Galactic globular clusters. We center our attention here in
  the preliminary analysis of three of these clusters: NGC~6441,
  Terzan~10 and 2MASS-GC02.
\end{abstract}

\section{Introduction}
The globular cluster (GC) system in the Milky Way is formed by around
150 known objects. The Galactic GCs are old systems, representing a
fossil record of the Galaxy, and can provide us with interesting hints
about its formation and its early evolution. As such, a proper
understanding of the Galactic GCs physical parameters is of
considerable importance. Photometrically, the tools more used to study
the Galactic GCs and get reliable values for their distances,
metallicities, ages, extinctions and other characteristics, have been
the analysis of their color-magnitude diagrams (CMDs) and the analysis
of the light-curves of their member variable stars. The most common
and characteristic variable type of star in GC is the RR~Lyrae type.
These variables lie in the instability strip region that crosses the
horizontal branch (HB) in the CMD of the cluster. RR~Lyrae in Galactic
GCs are classified in two main types according to the radial mode in
which they pulsate: RRab (or RR0) pulsate in the fundamental radial
mode, and RRc (or RR1) pulsate in the first overtone radial mode.  An
interesting characteristic of Galactic GCs in which RR~Lyrae variables
are present, is the so-called Oosterhoff dichotomy \citep{cat09,sm11}:
the Galactic GCs tend toward clumping in two main groups, Oosterhoff I
with shorter fundamental periods ($<P_{ab}>\sim0.55$ days) and
Oosterhoff II with longer fundamental periods ($<P_{ab}>\sim0.64$
days), leaving an almost empty gap at $<P_{ab}>\sim0.60$ days. This
dichotomy seems not to be present in the GC system of our neighbor
galaxies.

Most of the Galactic GCs lie inside the Solar Circle, and a
significant number of them are located towards the Galactic
center. Observations towards low Galactic latitudes where most of the
inner Galactic GC lie, are highly complicated by the considerable
amount of extinction produced by the gas and dust present. Therefore
the study of these GCs have been historically neglected, and their
physical parameters have not been as accurately determined as the ones
from the GCs located in other regions with low or negligible
extinction \citep{al12}. This situation is rapidly changing with the
advent of new state-of-the art telescopes and instruments to observe
at near-infrared wavelengths where extinction is highly diminished
($A_{Ks}\sim0.1A_V$). One project that will greatly increase our
knowledge of the inner Galactic GCs is the VVV survey.

\section{VVV survey}
The {\it Vista Variables in the Via Lactea} (VVV) survey is a
currently ongoing ESO public survey \citep{mi10,sa10,cat11}. It is
being conducted with the 4m Vista telescope located in the Cerro
Paranal Observatory, in Chile. The Vista Telescope, and VIRCAM, its
wide-field camera, are especially suited for near-infrared
observations. The VIRCAM camera provides observations with a field of
view of $1.48\times1.11$ deg$^2$ and a spatial resolution of $0''.34$
per pixel. The VVV is surveying 562 deg$^2$ in the Galactic bulge
($-10.0^{\circ}<l<+10.5^{\circ}$ and $-10.3^{\circ}<b<+5.1^{\circ}$ )
and in an adjacent region of the Galactic disc
($-65.3^{\circ}<l<-10.0^{\circ}$ and
$-2.25^{\circ}<b<+2.25^{\circ}$). A total of 36 known globular
clusters lie in the region surveyed \citep{ha96,mi10}, plus new GC
candidates are being discovered \citep{mi11b,mo11}. The survey is
going to provide an atlas of the Galactic inner regions in five
near-infrared filters ($Z$, $Y$, $J$, $H$, and $K_s$), and the
light-curves in $K_s$ for the objects found. Observations in the
different filters have already been taken, along with observations in
$K_s$ in some different epochs, enough to produce preliminary
CMDs and light-curves for the stars in the surveyed inner Galactic
GCs. In the next sections we will focus in three of them: NGC~6441,
Terzan~10 and 2MASS-GC02.
 
\section{NGC~6441}
NGC~6441 is not a common metal-rich globular cluster. In addition of
having a short red HB, typical of metal-rich globular clusters, it has
a blue HB component, as first found by \citet{ri97}.  It also presents
a significant amount of variables, and among them a high number of
RR~Lyrae \citep{la99,pr01,co06}, another rare characteristic in
metal-rich clusters. Moreover these variables have unusually long
periods, which has conducted to classify this GC, along with its twin,
NGC~6388, in a completely new Oosterhoff type: the Oosterhoff III
class.  Differential reddening across its field makes its optical CMD
difficult to interpret \citep{la99,pr01}. But the different
evolutionary sequences get narrower in the near-infrared CMD (see
figure \ref{figcmd6441}), allowing the comparison with theoretical
models. In figure \ref{figcmd6441} we have overplotted an old-age
isochrone from the PGPUC library \citep{va12}, the only one so far
that provides isochrones in the Vista filter system. Using the values
from iron content, distance, and color excess from the Harris catalog
\citep{ha96}, we can see that the agreement with the observational CMD
is good. Unfortunately, our data does not allow us to accurately
define the main-sequence turn-off point, and that way better constrain
the age of the cluster.

\articlefigure[scale=0.35]{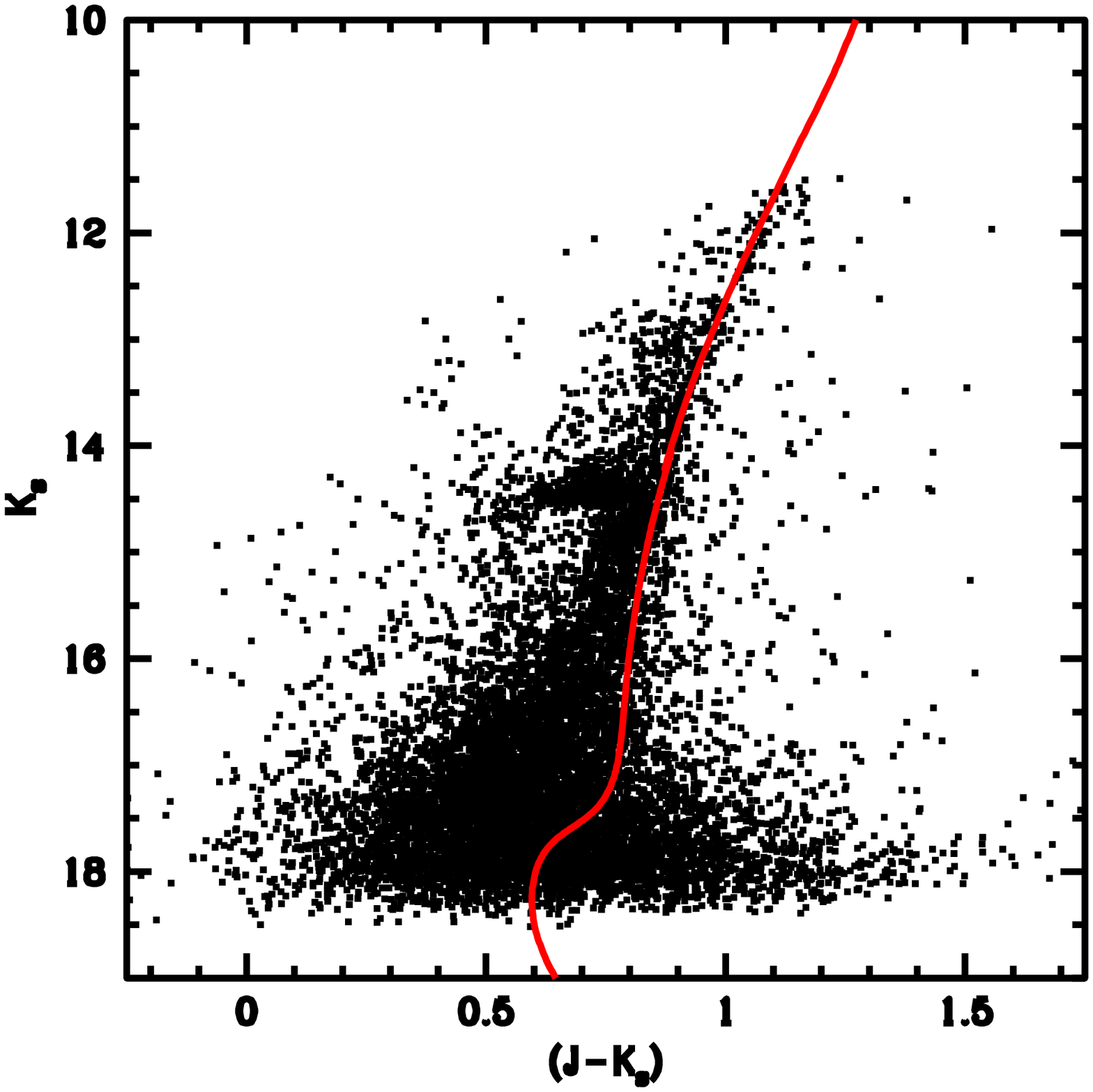}{figcmd6441}{Near-infrared
  CMD of the inner $2'$ of NGC~6441. We have overplotted a 13~Gyr
  isochrone from the PGPUC library \citep{va12}, with the iron content
  and moved using the distance and color-excess provided in the
  current version of the Harris catalog \citep{ha96}. }
\articlefigure[scale=0.35]{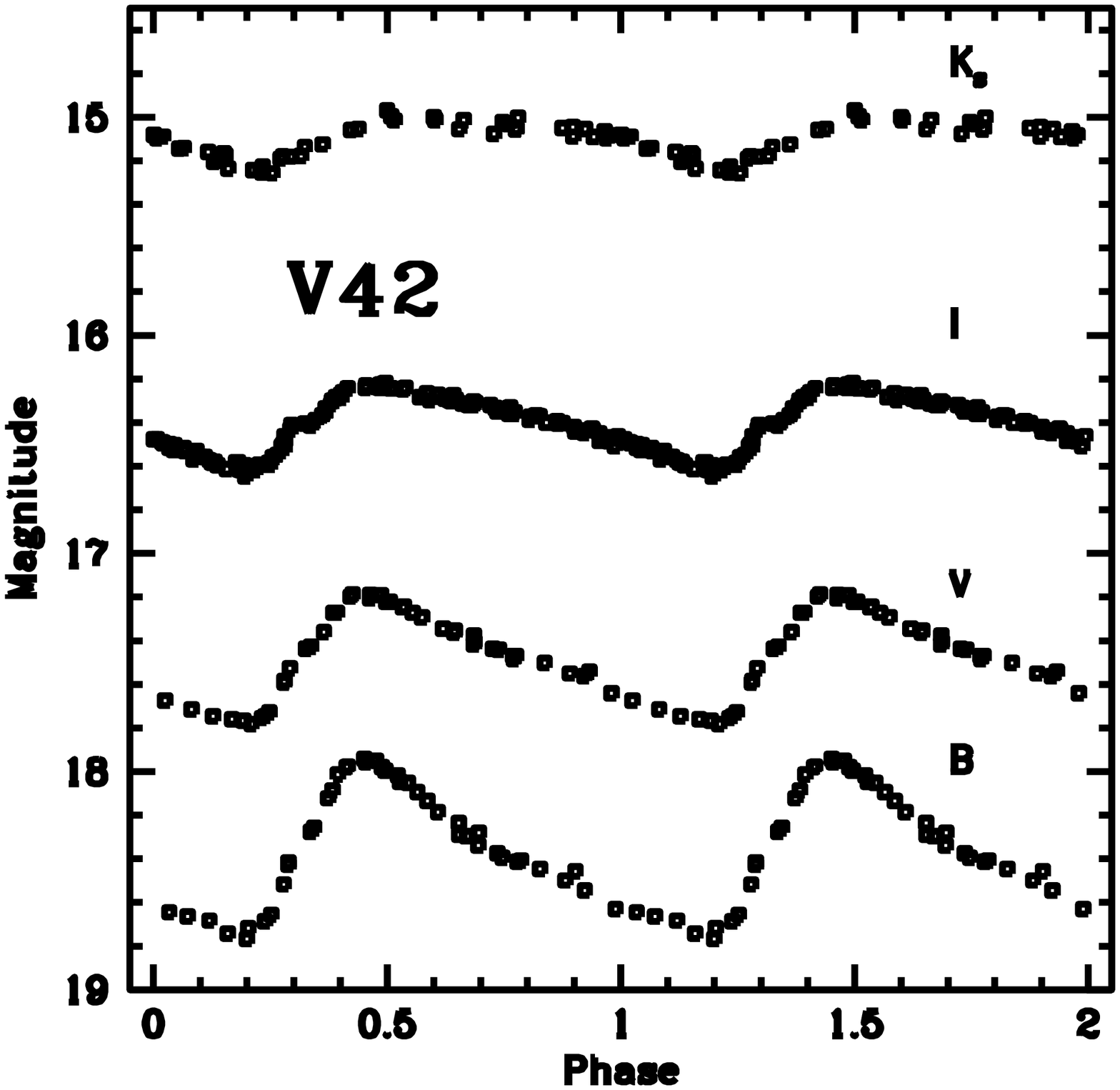}{figcurve6441}{Light-curves
  at different wavelengths for one of the already-known NGC~6441
  variables, V42, a RRab type star. The $K_S$ light-curve comes from
  our VVV data, while the I light-curve comes from the OGLE survey
  \citep{so11}, and the B and V data from the study by \citet{pr01}. }

There are more than 150 variables already discovered in this GC, many
of which are RR~Lyrae. Although at near-infrared wavelengths their
amplitude is reduced and the shape of the RRab light-curves is more
sinusoidal (see figure \ref{figcurve6441}), the quality of our data
and the present sample of the light-curve (45 epochs for NGC~6441)
allow us to look for more variables, and to characterize amplitudes,
periods and mean magnitudes and colors. Using these parameters and the
RR~Lyrae period-luminosity relation in the near-infrared
\citep{bo01,cas04,cat04}, we will better define the extinction and
distance to this GC.

\section{Terzan~10 and 2MASS-GC02}
Terzan~10 and 2MASS-GC02 are recently discovered, highly-extincted,
low-mass, Galactic globular clusters (see Harris
catalog). Terzan~10 was discovered by \citet{te71}, and
2MASS-GC02, even more recently, by \citet{hu00}, using data from
2MASS. The high extinction in their line of sight ($A_V\sim7.5$ for
Terzan~10, and $A_V\sim16$ for 2MASS-GC02) makes them extremely faint
and difficult to observe in the optical bands, but not at
near-infrared bands.

The VVV data allow us to produce CMDs for these objects from the tip
of the RGB down to the SGB region (see figure \ref{figcmd2}). But the
analysis of the CMD here is complicated by other fact: their low
number of member stars, and the high number of field stars in the
regions they are located. In addition, differential reddening across
the field of both GCs is clearly present in their near-infrared CMDs,
despite reddening being highly diminished at these wavelengths, making
their CMDs even more difficult to interpret.

In both cases the presence of RR~Lyrae would help to better constrain
these GC parameters. \citet{bo07} found five RR~Lyrae candidates for
2MASS-GC02, but no candidates have been reported so far for
Terzan~10. We have not been able to confirm any of the candidates
found in \citet{bo07}, but we have found 5 more RR~Lyrae candidates in
2MASS-GC02, and for the first time, we have found 5 RR~Lyrae
candidates in Terzan~10 (see figure \ref{figlc2}). Our selection of
RR~Lyrae candidates was done according to shape of the light-curve,
period, and position in the CMD (see figures \ref{figcmd2} and
\ref{figlc2}). At present, we have 33 epochs in 2MASS-GC02, which more
than doubles the number of epochs in \citet{bo07} study, and 90 epochs
in Terzan~10. A surprising fact in these two GCs is their Oosterhoff
classification. 2MASS-GC02 has a mean period for their RR~Lyrae
candidates of $<P>\sim0.61$, which puts it in the Oosterhoff gap,
while Terzan~10 has a mean period for their RR~Lyrae candidates of
$<P>\sim0.65$, making it an Oosterhoff II candidate, which is strange
for the reported metallicity of this object, $[Fe/H]=-1.00$ according
to the Harris catalog \citep{ha96}. These results, however, should be
considered with care due to the small number of RR~Lyrae candidates
found so far in these two objects. We should also note that the
reported metallicities from literature do not come from
high-resolution spectroscopy. We are currently refining our analysis
to try to find more RR~Lyrae variables in these GCs and provide a more
complete and statistically significant sample of variables for these
two GCs.

\articlefiguretwo{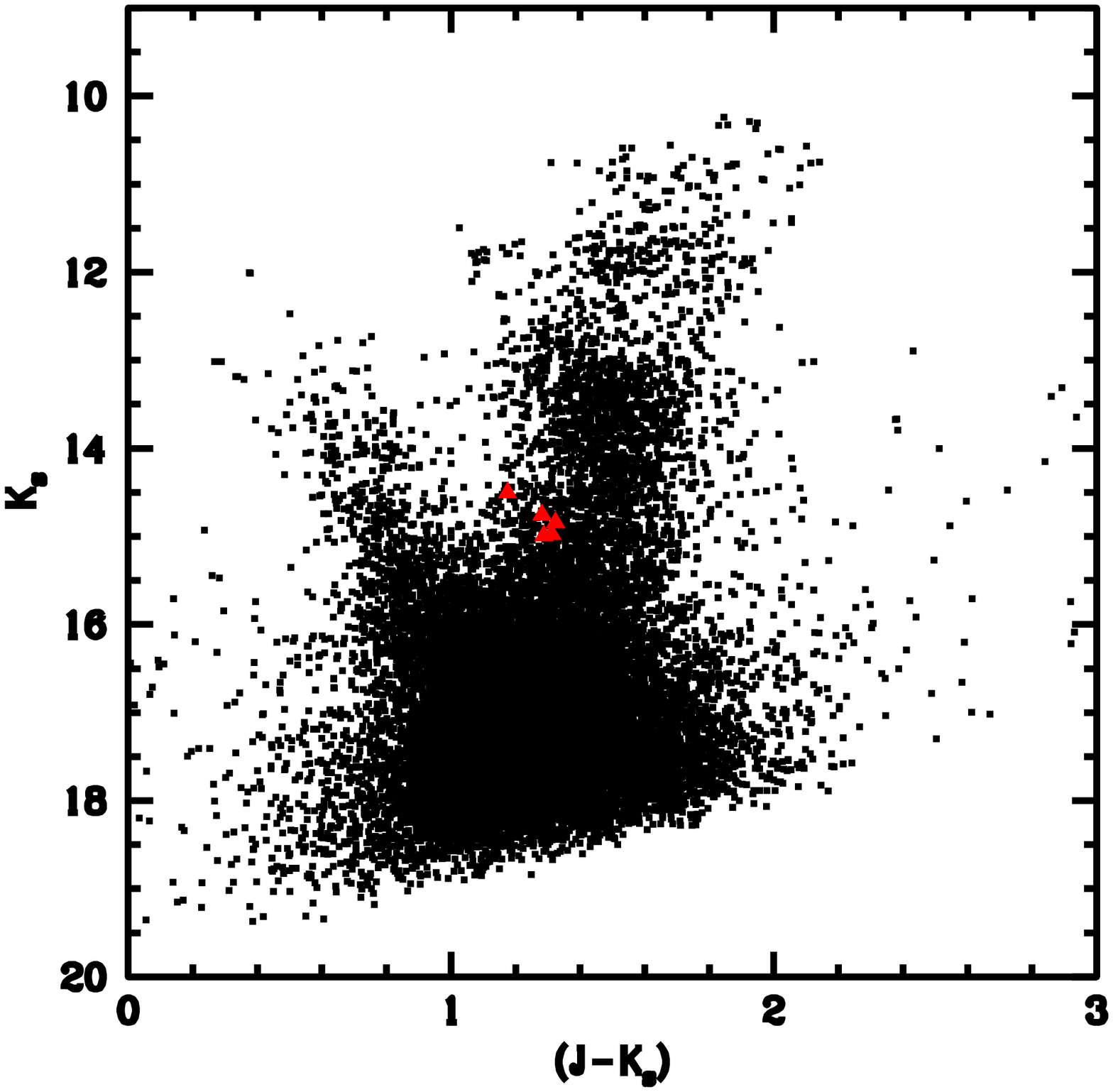}{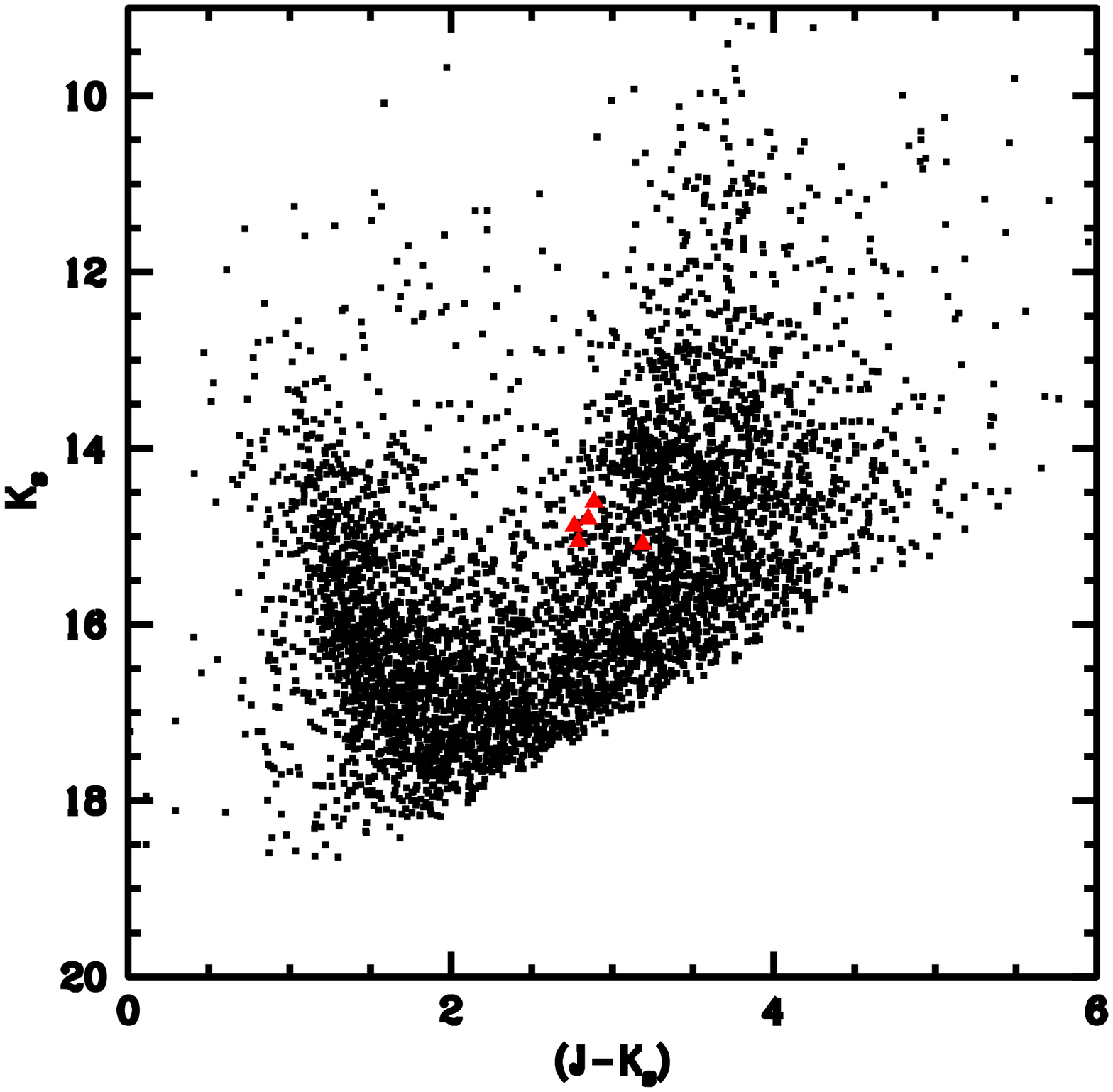}{figcmd2}
{Near-infrared CMDs of the inner 2' of Terzan~10, on the left, and of
  2MASS-GC02, on the right. Our RR~Lyrae candidates are marked as red
  triangles.}
\articlefiguretwo{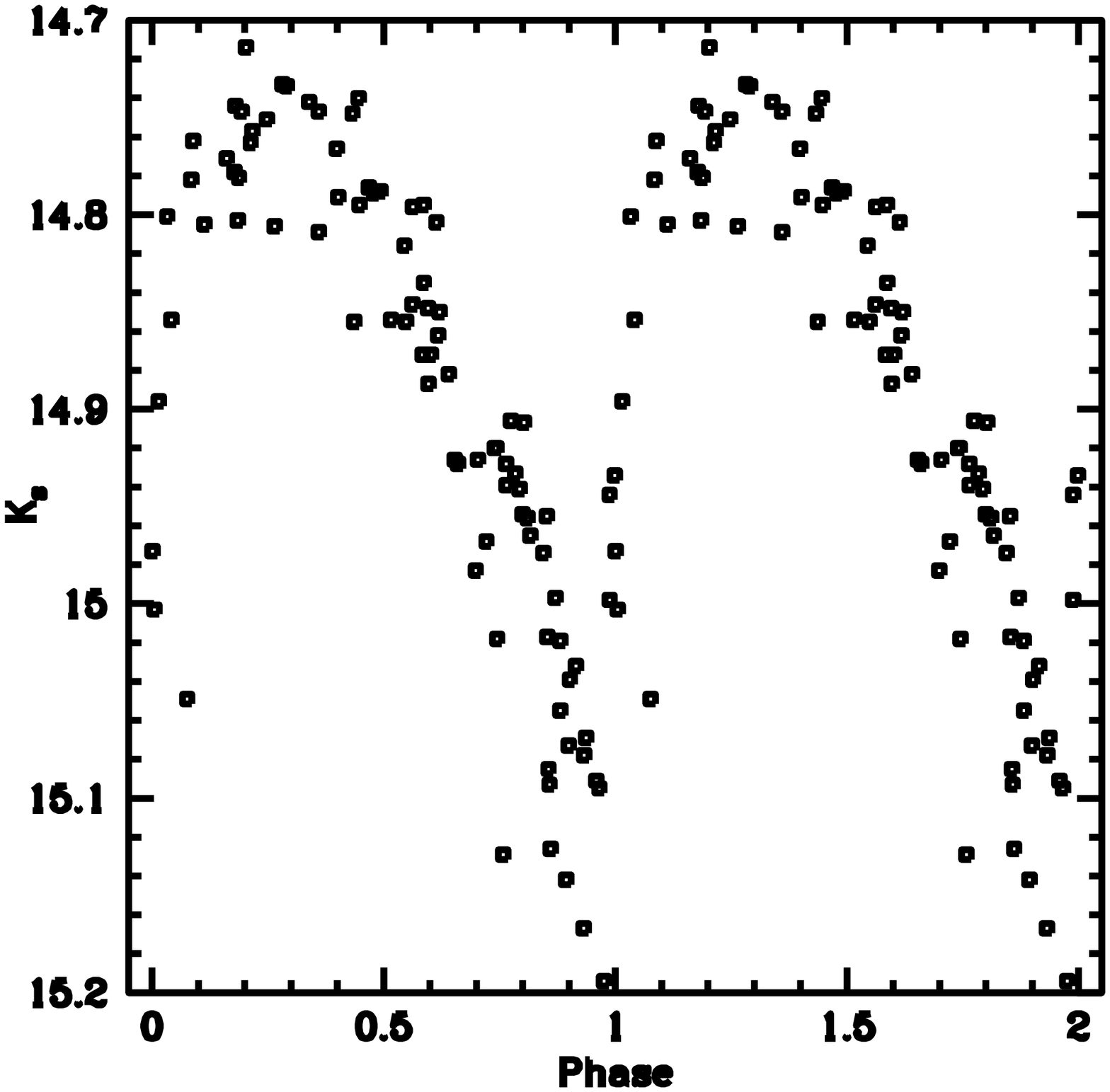}{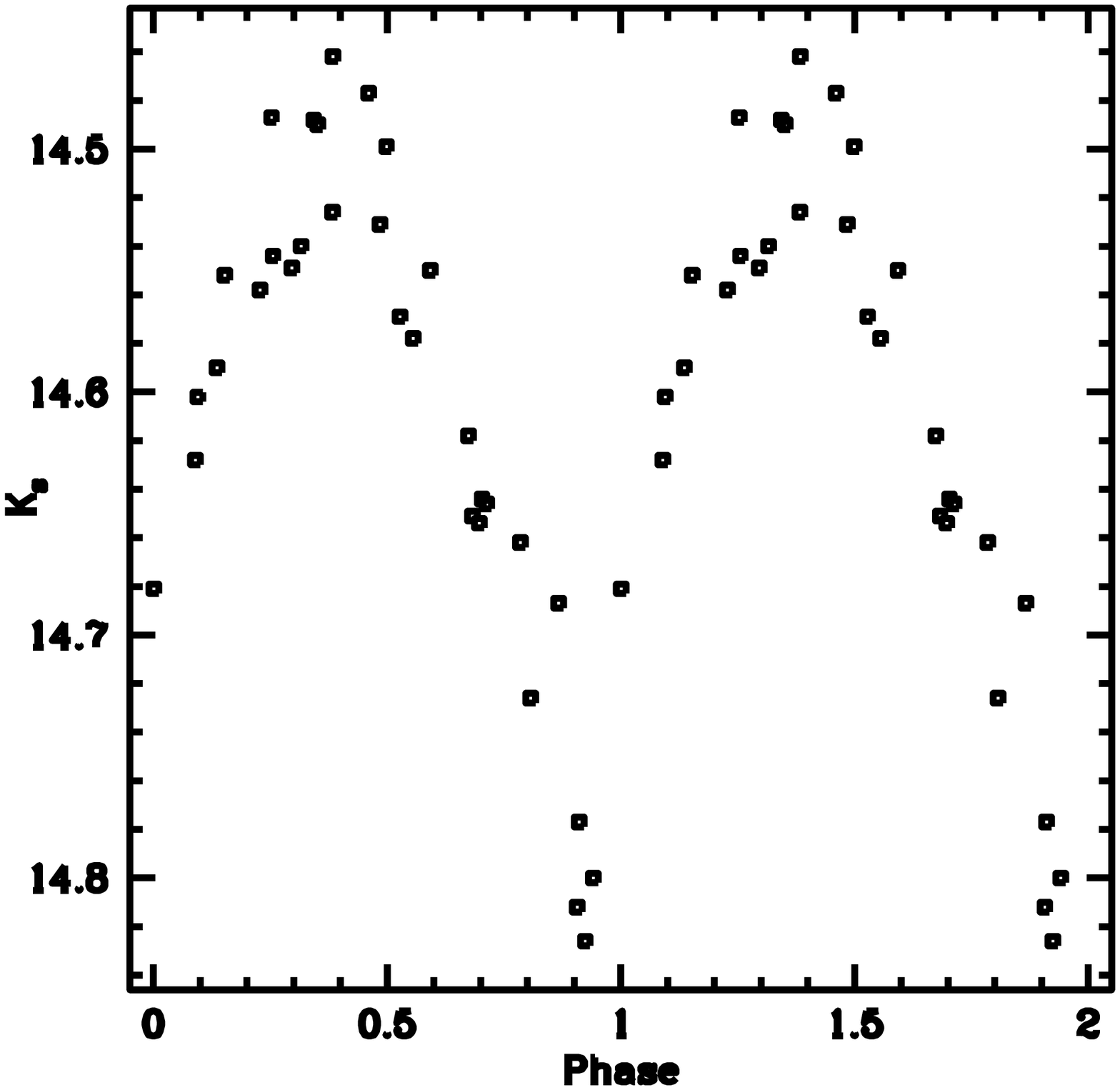}{figlc2}
{Light-curves of one of the new RR~Lyrae candidates found in
  Terzan~10, on the left, and in 2MASS-GC02, on the right. Note that
  the current number of epochs in the Terzan10 candidate is almost
  three times the number in 2MASS-GC02, but at the end of the survey
  the number will be the same.}

\section{Conclusions}
The VVV survey will provide a highly improved view of the GCs located
in the inner regions of the Milky Way. The use of near-infrared
wavelengths will allow us to penetrate through the gas and dust that
hide these GCs and produced clean CMDs and high-quality light-curves
for their variable members. Analysis of those CMDs and light-curves
will let us infer more accurately these objects physical parameters
and finally include the precise characteristics of this elusive sample
in the studies of GC in our Galaxy. In these proceedings, and as
examples of the coming analysis, we have shown the CMDs and the
RR~Lyrae light-curves for three GCs that lie inside the region
surveyed: NGC 6441, Terzan~10 and 2MASS-GC02. A preliminary analysis
from its CMD show that NGC~6441 values for its distance and extinction
agree quite well with current values from literature, while the CMDs
from Terzan~10 and 2MASS-GC02 are more difficult to analyze. For these
last two clusters, we have found new RR~Lyrae variable
candidates. Their Oosterhoff classification suffers from small
statistics, but looks rather uncommon. Terzan~10 seems to be an
Oosterhoff II type but with too high an iron content to belong to the
group, while 2MASS-GC02 seems to be in the Oosterhoff gap, where very
few Galactic GC lie.
  
\acknowledgements This project is supported by the Chilean Ministry
for the Economy, Development, and Tourism's Programa Iniciativa
Cient\'ifica Milenio through grant P07-021-F, awarded to The Milky Way
Millennium Nucleus; by Proyecto Fondecyt Postdoctoral 3130552; by
Proyecto Fondecyt Regular 1110326; by Proyecto Basal CATA PFB-06; and
by Anillos ACT-86.

\end{document}